\documentclass[sigconf]{acmart}
\acmConference[ICSE 2024]{46th International Conference on Software Engineering}{April 2024}{Lisbon, Portugal}

\usepackage[font=small]{caption}
\usepackage{lipsum}
\usepackage{natbib}
\usepackage{comment}
\usepackage{booktabs}
\usepackage{pbox}
\usepackage{amsmath}
 
\usepackage{tablefootnote}
\usepackage{gensymb}
\usepackage{epsfig}
\usepackage{amssymb,amsfonts,latexsym}
\usepackage{enumerate}
\usepackage{xspace}
\usepackage{epsf,picinpar}
\usepackage{varioref}
\usepackage{colortbl,multirow,hhline}
\usepackage{amssymb}
\usepackage{colortbl,multirow,hhline}
\usepackage{algorithmic}
\usepackage{algorithm}
\usepackage[normalem]{ulem}
\usepackage{pifont}
\usepackage{colortbl}
\usepackage{url}
\usepackage{balance}
\usepackage{graphicx}
\usepackage{longtable}
\usepackage{lscape}
\usepackage{multirow}
\usepackage{framed}
\usepackage{morefloats}
\usepackage[T1]{fontenc}
\usepackage{array}
\usepackage{pdfpages}
\usepackage{fancybox}
\usepackage{adjustbox}
\usepackage{flushend}
\usepackage{booktabs}
\usepackage{enumitem}
\usepackage{tabularx}
\usepackage{diagbox}
\usepackage{soul} % highlighting
\usepackage{pgfplots} % Bar chart
\usepackage{booktabs, multirow} % for borders and merged ranges
\usepackage{soul}% for underlines
\usepackage{changepage,threeparttable} % for wide tables
\usepackage{placeins} % Forçar Tabela/Figura fica abaixo do texto
\usepackage[most]{tcolorbox}
\usepackage{graphicx}
\usepackage{longtable}
\usepackage{dcolumn}
\usepackage{xltabular}
\usepackage{soul}
\usepackage{subfig}
\usepackage{color}
\usepackage{geometry}
\usetikzlibrary{mindmap,trees,backgrounds,shadows}

\makeatletter
\newcommand\BeraMonottfamily{%
  \def\fvm@Scale{0.75}% scales the font down
  \fontfamily{fvm}\selectfont% selects the Bera Mono font
}
\makeatother

\usepackage{soul}% http://ctan.org/pkg/soul
\usepackage{marvosym} % icons

\newcommand{\survey}{62\xspace}
\newcommand{\tool}{StackSpot AI\xspace}
\newcommand{\zup}{\textsc{Zup Innovation}\xspace}

\AtBeginDocument{%
  \providecommand\BibTeX{{%
    \normalfont B\kern-0.5em{\scshape i\kern-0.25em b}\kern-0.8em\TeX}}}

\pgfplotsset{compat=1.18}

\begin{document}

\title{Developer Experiences with a Contextualized AI Coding Assistant: \\ Usability, Expectations, and Outcomes}

\author{Gustavo Pinto}
\affiliation{%
  \institution{Zup Innovation \& UFPA}
  \city{Belém}
  \state{PA}
  \country{Brazil}
}
\email{gustavo.pinto@zup.com.br}

\author{Cleidson de Souza}
\affiliation{%
  \institution{UFPA}
  \city{Belém}
  \state{PA}
  \country{Brazil}
}
\email{cleidson.desouza@acm.org}

\author{Thayssa Rocha}
\affiliation{%
  \institution{Zup Innovation \& UFPA}
  \city{Belém}
  \state{PA}
  \country{Brazil}
}
\email{thayssa.rocha@zup.com.br}

\author{Igor Steinmacher}
\affiliation{%
  \institution{Northern Arizona University}
  \city{FlagStaff}
  \state{AZ}
  \country{EUA}
}
\email{Igor.Steinmacher@nau.edu}

\author{Alberto de Souza}
\affiliation{%
  \institution{Zup Innovation}
  \city{São Paulo}
  \state{SP}
  \country{Brazil}
}
\email{alberto.tavares@zup.com.br}

\author{Edward Monteiro}
\affiliation{%
  \institution{StackSpot}
  \city{São Paulo}
  \state{SP}
  \country{Brazil}
}
\email{edward.monteiro@stackspot.com}

%%
%% The ``author'' command and its associated commands are used to define
%% ``authornote'' and ``authornotemark'' commands



%%%%%%%%%%%%%%%%%%%%%%%%%%%%
% remove copyright for submission
%\settopmatter{printacmref=false}
%\setcopyright{none}
%\renewcommand\footnotetextcopyrightpermission[1]{}
%\pagestyle{plain}
% page numbering
%\settopmatter{printfolios=true} 
%%%%%%%%%%%%%%%%%%%%%%%%%%%%

\begin{abstract}
In the rapidly advancing field of artificial intelligence, software development has emerged as a key area of innovation. Despite the plethora of \emph{general-purpose} AI assistants available, their effectiveness diminishes in complex, domain-specific scenarios. Noting this limitation, both the academic community and industry players are relying on \emph{contextualized} coding AI assistants. These assistants surpass general-purpose AI tools by integrating proprietary, domain-specific knowledge, offering precise and relevant solutions. Our study focuses on the initial experiences of \survey participants who used a contextualized coding AI assistant --- named \tool--- in a controlled setting. According to the participants, the assistants' use resulted in significant time savings, easier access to documentation, and the generation of accurate codes for internal APIs. However, challenges associated with the knowledge sources necessary to make the coding assistant access more contextual information as well as variable responses and limitations in handling complex codes were observed. The study's findings, detailing both the benefits and challenges of contextualized AI assistants, underscore their potential to revolutionize software development practices, while also highlighting areas for further refinement.
\end{abstract}


\maketitle

\section{Introduction}

In the rapidly evolving landscape of artificial intelligence (AI) and its integration into various industries, the software development domain remains at the forefront of innovation~\cite{fan2023large}. Developers today are equipped with an unprecedented set of coding tools and AI agents, designed to navigate and simplify the complexities of software engineering. However, as software projects grow intricate, the demand for precise and efficient coding assistance becomes critical~\cite{chen2017codeon}.

Despite their groundbreaking nature, general-purpose AI assistants currently have a fundamental shortcoming: they often provide generic or inaccurate responses, particularly when confronted with contextualized, domain-specific queries \cite{chan2023chateval}. 
This gap is felt by developers who seek guidance, for example, in tasks related to optimizing a database query~\cite{Anonymous} or deciphering the complexities of a proprietary codebase. Conventional AI tools, while advanced in many aspects, often fall short of delivering the depth and specificity required in these scenarios. Such limitations not only hinder productivity but also pose a barrier to harnessing the full potential of AI in software development.

The industry is responding to these challenges by developing contextualized coding AI assistants. These tools, underpinned by advanced AI models, are specifically designed to access and utilize proprietary, domain-specific knowledge, which general-purpose assistants typically lack. This specialized approach enables them to offer targeted assistance, especially useful in complex, domain-specific scenarios. To illustrate, imagine a scenario where a developer is working on an intricate e-commerce platform, and they encounter a challenge related to optimizing a multi-tier product recommendation algorithm. While a general-purpose AI might offer broad-based guidance or algorithmic solutions, a contextualized coding assistant, familiar with the proprietary nuances of that specific e-commerce platform and its surroundings, might pinpoint exact issues based on historical data or even provide solutions that account for platform-specific constraints, or company-based frameworks. In essence, while a general-purpose AI tool might suggest generic best practices, a contextualized assistant could reference company-specific requirements documents and related projects, offering answers that are not only effective but also tailored to that organization's unique needs. For instance, Enterprise Tabnine customers might choose to train their own custom model based on their company's source code\footnote{\url{https://www.tabnine.com/code-privacy}}.

%Example of Contextualized Coding AI Assistants include \xxx, \xxx, and \xxx. 
These contextualized tools, while differing in purpose from general-purpose coding assistants, often share a common technological foundation: they leverage the Retrieval-Augmented Generation (RAG) technique~\cite{Lewis:NIPS:2020}. This involves retrieving relevant information from specialized sources and generating contextualized responses using advanced language models. An example of this is the contextualized coding AI assistant name \tool developed by \zup\footnote{\tool and \zup are two pseudonyms adopted for double anonymous purposes.}--- a software partner tech company --- aimed at enhancing developer productivity, confidence, and experience with AI-based tools. The main capabilities of this assistant are detailed in Section~\ref{sec:stk-ai}. %\textcolor{red}{Podemos falar isto? a submissão é anônima.}

In this paper, we report the findings from a study about the experience of \survey practitioners who used the \tool assistant for the first time in a controlled, online environment. %Most of these practitioners were not familiar with coding AI assistants (neither general nor contextualized). 
During four hours, they were introduced to concepts and usage details of \tool, followed by a hands-on experience performing simple tasks. The participants interacted and provided feedback during the whole online discussion. The analysis of their feedback highlights several key [\textbf{B}]enefits and [\textbf{C}]hallenges encountered:
% They were also requested to perform additional tasks and provide feedback during an online discussion. The analysis of their feedback highlights several key [\textbf{B}]enefits and [\textbf{C}]hallenges encountered:

\begin{itemize}
    \item[\textbf{[B]}] Generation of accurate codes for swift integration with internal APIs and support for routine tasks;
    \item[\textbf{[B]}] Time efficiency by centralizing information access;
    \item[\textbf{[B]}] Streamlined access to documentation and guidelines within the IDE;
    \item[\textbf{[C]}] Multiple knowledge sources are required to maintain response accuracy;
    \item[\textbf{[C]}] Inconsistency in responses to identical prompts, requiring prompt refinement; and
    \item[\textbf{[C]}] Difficulties in generating complex code structures.% like ready controllers.
    %Limit the size of the context sent to the AI.
\end{itemize}

%\begin{itemize}
%    \item[\textbf{[B]}] Time savings by not having to search for information in several different places.
%    \item[\textbf{[B]}] Easier access to documentation and guidelines within the IDE environment.
%    \item[\textbf{[B]}] Generation of accurate and complete codes to swiftly integrate with internal APIs. Support for repetitive and routine tasks through "quick commands".
%    \item[\textbf{[C]}] Variation in responses to the same prompt. Need to refine and adjust prompts to obtain accurate responses.
%    \item[\textbf{[C]}] Difficulty in generating more complex codes like ready controllers.
%    \item[\textbf{[C]}] Mixing of different knowledge sources affecting accuracy. Limit the size of the context sent to the AI.
%\end{itemize}

Furthermore, the participants provided important insights, bringing feedback on the experience, suggestions on new functionalities, and reflections on the productivity and reliability of code generated by \tool.

The rest of the paper is organized as follows. Section ~\ref{sec:related} presents related work on AI coding assistants' usage. Section ~\ref{sec:stk-ai} presents the built tool and its value proposal. Section ~\ref{sec:study} outlines the experiment details, followed by Section ~\ref{sec:results} that presents the obtained results. In Section ~\ref{sec:discussion} we engage in discussion on our findings, while Section ~\ref{sec:limitations} discusses some limitations of this study. Section ~\ref{sec:conclusion} brings conclusions and next steps for research and evolution of \tool.

\section{Related Work}
\label{sec:related}

Research in AI assistants has focused on different aspects including the benchmarks necessary to evaluate and compare them~\cite{chen2021evaluating}, the correctness~\cite{ Yetistiren2022}, complexity ~\cite{Nguyen2022}, quality~\cite{Dakhel2023}, and security~\cite{pearce2022} of the generated code, the developers' experience while using these assistants~\cite{vaithilingam2022expectation, barke2023grounded}, among other aspects. In this paper, we are interested in two aspects. First, the user experience using these tools. Second, the correctness of the solutions generated %by these tools%
, i.e., their ability to, given a particular problem, generate a code solution that will actually solve that problem. This is measured by checking whether the solution passes the test cases associated with the original problem. 

\subsection{User Experience}

We can find a few papers discussing the user experience of software developers using AI code assistants including \cite{Xu2022,jiang2022discovering, Bird2023, vaithilingam2022expectation, barke2023grounded, Anonymous}. In general, these studies indicate that developers save time using AI assistants, i.e., \textit{``interactions with programming assistants are bimodal: in acceleration mode, the programmer knows what to do next and uses Copilot to get there faster; in exploration mode, the programmer is unsure how to proceed and uses Copilot to explore their options''}~\cite{barke2023grounded}. Even when the assistants are not 100\% correct, they still generate code that can be used as a ``starting point'' for further work.

These studies also reported some of the limitations of these tools, mainly lack of correctness of the code suggestions and interruptions, i.e., the assistants disturb the natural flow of work~\cite{Bird2023, Anonymous}. More interestingly, they report coping strategies to deal with Copilot’s limitations: \textit{``to accept the incorrect suggestion and attempt to repair it,''} add more context so that the assistant improves its suggestions, or simply stop using the tool.

\subsection{Correctness}
In 2022, two different papers were published assessing the correctness of GitHub Copilot. In the first paper, Nguyen and Nadi~\cite{Nguyen2022} assessed the correctness of Copilot's suggestions in four different programming languages: Java, JavaScript, Python, and C. Each programming language had a different result with Python code generated by Copilot with a 42\% correctness, while Java had 57\% and JavaScript with 27\%. These authors tested the code generation abilities to solve 33 questions randomly selected from LeetCode,  a popular Question Pool website with several various coding questions on different topics (array, algorithm, sorting, etc). 

Meanwhile, Yetistiren and colleagues~\cite{Yetistiren2022} focused solely on Python and used the HumanEval~\cite{chen2021evaluating} benchmark, the same one used to evaluate Codex, the GPT model behind Copilot. This benchmark contains 164 \textit{original} programming problems "with some comparable to simple software interview questions". In their result, Copilot's suggestions had a 28.7\% correctness rate.

Several factors might explain the different correctness rates in these studies (42\% vs 27.8\%). Arguably, a potential explanation is associated with the datasets used. While Nguyen and Nadi~\cite{Nguyen2022} used a \textit{popular} programming site, Yetistiren et al.~\cite{Yetistiren2022} used \textit{original} programming problems, i.e., a popular programming site like LeetCode might even be used in the Copilot's training dataset. This seems to suggest that Copilot's correctness is influenced by the presence of similar data in its training dataset. Therefore, when faced with domain-specific queries, Copilot is likely to provide generic or inaccurate suggestions.

\section{\tool}\label{sec:stk-ai}

In this section, we describe how \tool works.

\subsection{Approach}
\label{sec:approach}

Different than Copilot or CodeWhisperer, which are \emph{general-purpose} coding assistants, \tool is a highly \emph{contextualized} coding AI assistant. \tool takes into account the nuanced requirements of individual developers and the intricacies of specific projects (codebases). This tailored approach is based on the implementation of the Retrieval Augmented Generation (RAG) mechanism~\cite{Lewis:NIPS:2020}. 

RAG is an approach designed to enhance LLM-generated content by anchoring it in external knowledge sources. In question-answering systems, RAG accesses up-to-date, reliable information and provides transparency to users regarding the model's information sources, promoting trust and verifiability. So, this approach mitigates the risk of sensitive data leakage and misinformation generation while also improving response quality. An illustrative list of possible knowledge sources includes:

\begin{enumerate}
    \item An extensive catalog of APIs recurrently harnessed by the development team;
    \item Exemplary code snippets serving for discerning coding paradigms or facilitating code modernization activities;
    \item Customized artifacts written in natural language, including but not limited to, guidelines delineating the protocol for repository commits and a comprehensive list of software requirements to be implemented. 
\end{enumerate}

By providing relevant and up-to-date information to the LLM, RAG also reduces the need for constant model retraining and parameter updates, lowering computational and financial overhead~\cite{Lewis:NIPS:2020}, since there is no need to build a new foundation model or retrain an existing one. Additionally, RAG is a two-pronged structure consisting of ``retrieval'' and ``generation'' components~\cite{Lewis:NIPS:2020}.

The ``retrieval'' facet of RAG is designed to fetch relevant documents from a specified dataset. Traditional databases might falter in efficiently retrieving relevant documents. In \tool, we use information retrieval techniques to identify the most relevant document for a given user query. Although the retrieval component is efficient at sourcing relevant information, it does not have the capability to generate new content.

On the flip side, the ``generation'' component harnesses the prowess of OpenAI's most recent model, GPT-4. Imagine a scenario where a developer is conceptualizing a new algorithm but hits a roadblock in terms of its implementation. \tool, channeling the generative capabilities of GPT-4, can aid in generating code snippets that are tailored to the developer's specific context, based on the documents found by the retrieval component.

In essence, \tool joins advanced contextual retrieval with state-of-the-art generative capabilities, ensuring developers receive precise, contextual, and timely assistance. It does so by using OpenAI's newest model, GPT-4. %coloquei pois está implicito no texto, a idéia é deixar mais explicito.

\subsection{Prep-and-Go}
\tool has two main interfaces. The first one is a web portal in which users can configure their teams' preferences and upload representative documents, which would be later used by the retrieval component. These preferences' configurations allow the use of recommended development tech stacks and code patterns that are often employed in the development team. As such, the generated code might respect these stacks, minimizing the developer's effort in translating the generated response into their codebase.

Once the configuration is done, users can turn their attention to the \tool plugin that is currently available for VSCode and IntelliJ. Using this plugin, users could interact with the second interface: a coding assistant chatbot. In this way, developers can craft prompts, refine the answers, and copy the generated solution to the code editor. 
Additionally, it's important to note that \tool, utilizing GPT-4 as its core LLM, needs to manage the token limit imposed per request effectively. This token limit is crucial because exceeding it can lead to prompt overflow, a scenario where the number of tokens used exceeds the LLM's capacity. Given that \tool functions as a chatbot, it tracks and retains a history of the most recent messages exchanged with users. This historical data enriches the input prompt, enhancing the context and relevance of \tool's responses.
However, to prevent prompt overflow and maintain efficiency, \tool implements a strategy of selectively discarding older messages. This process ensures that the prompt remains within the token limit while retaining the most pertinent and recent interactions. Additionally, recognizing that users may shift topics during a conversation, \tool is designed to dynamically adjust which messages it retains. It prioritizes those that are most relevant to the current context of the dialogue. This adaptive approach ensures that \tool remains focused and relevant to the user's immediate needs, despite the evolving nature of the conversation.

\begin{figure*}[h]
    \centering
    \includegraphics[width=0.99\textwidth]{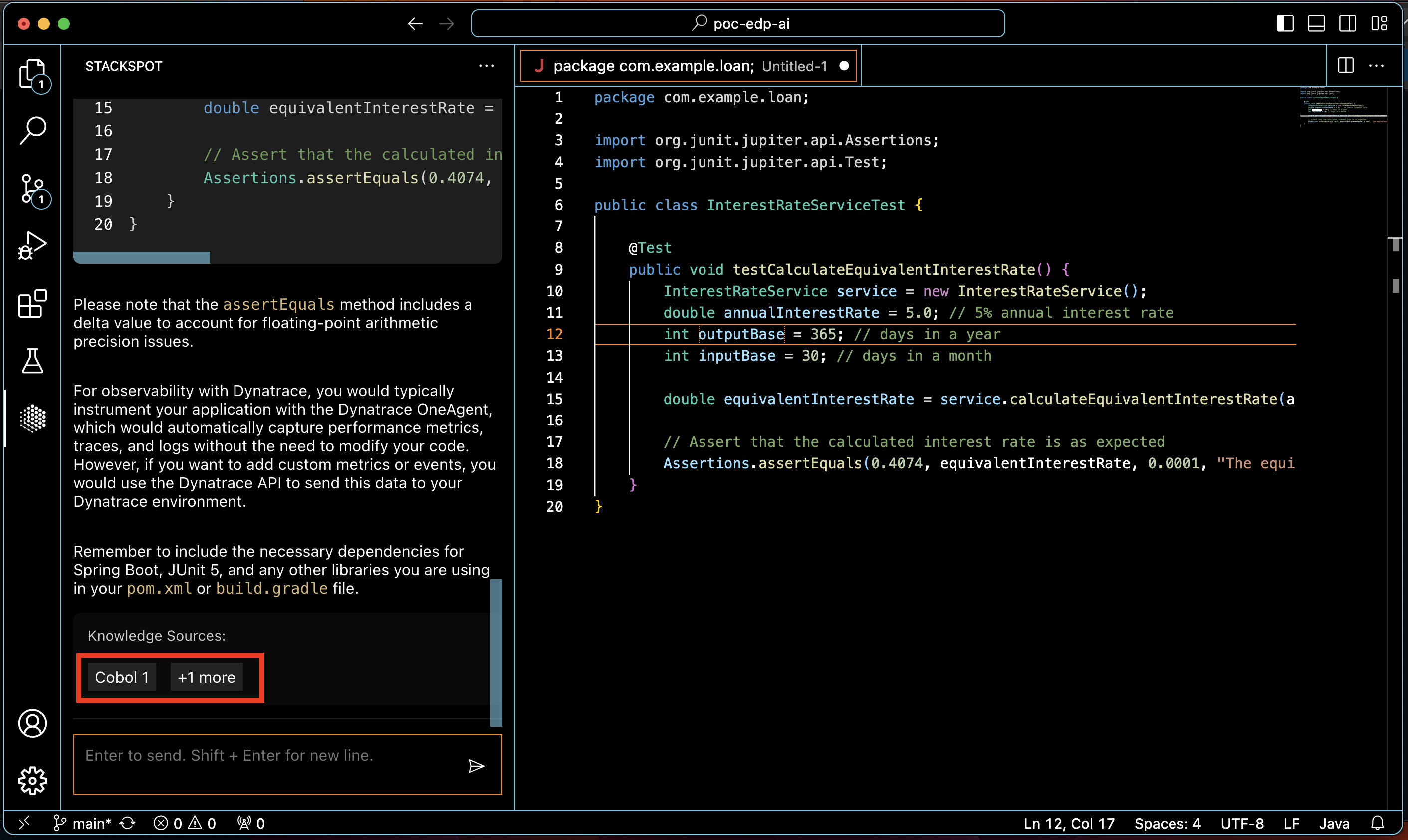}
    \caption{\tool plugin on VS Code.}
    \label{fig:plugin}
\end{figure*}

Figure~\ref{fig:plugin} shows an example of the use of the \tool plugin on VSCode. As one can see, \tool combines a code editor with a chat interface. The red box indicates the knowledge source found in the user search. 

% \tool has two main interfaces. It has a web portal in which users can configure their teams' preferences, but also upload representative documents, which would be later used by the retrieval component. After the configuration is done, users can turn their attention to the \tool plugin that is currently available for VSCode and IntelliJ. Using this plugin, users could interact with the coding assistant as a chatbot, crafting prompts, refining the answers, and copying the desired solution to the editor. Figure~\ref{fig:plugin} shows an example of the use of the \tool plugin on VSCode. As one can see, \tool combines a code editor with a chat interface. 

%\gnote{pensar se eu falo da arquitetura da \tool}

Finally, as a conversational agent, \tool extends its capabilities beyond mere generation of code snippets. It can engage users in broader discussions on various programming topics, offering insights and clarifications~\cite{Ross2023}. Additionally, it plays a role in enhancing users' programming skills through interactive learning and guidance, providing a more comprehensive, educational experience in the realm of software development.

\section{User Study}
\label{sec:study}
To evaluate the developers' experience in using \tool, we organized an in-company online study. The study was performed within \zup, a large software-producing company, with around 3.5k employees, and more than 10+ years in the market, working with some of the largest financial institutions in Latin America and abroad. For instance, for one of their clients, \zup engineers rewrote millions of Cobol legacy code into modern programming languages, helping to move their physical infrastructure to the cloud. In such a context, there is an important need for a text-based coding assistant, in particular, for modernization tasks.

The goal of the study was to introduce practitioners to \tool and gather representative feedback to improve the product's quality and usability. 
%We adopted an exploratory approach, with our primary focus being understanding the developer experience in using \tool. 
We intentionally refrained from setting specific design objectives for the tool, tailoring it to particular user groups (such as novices or experts) or specific scenarios (like coding or learning new programming languages), as we wanted any value provided by \tool to emerge from our user study.

In this section we describe how the study was planned, (Section~\ref{sec:studyplan}), how we collected data (Section~\ref{sec:data-collection}), and how we analyzed the data collected (Section~\ref{sec:data-analysis}). 

\subsection{Study Planning}\label{sec:studyplan}

The study was structured to provide participants with a journey into the capabilities and functionalities of \tool. Through an online meeting, facilitators stepped from introducing concepts and product demonstrations to hands-on exploration and collaborative discussions. The intention was to provide a comprehensive understanding of \tool. To do so, the planned timeline for the meeting was:

% The study was structured to provide participants with a journey into the capabilities and functionalities of \tool. From introducing concepts and product demonstrations to hands-on exploration and collaborative discussions, the agenda aimed to provide a comprehensive understanding of \tool. Here is a breakdown of the study:

\begin{itemize}[leftmargin=*]
    \item \textbf{Welcoming Session (2:00 PM – 2:15 PM)}: An initial moment to introduce ourselves and encourage those who did not answer the pre-study questionnaire to fill it out (more details about this survey are available in Section~\ref{sec:data-collection}).
    %An initial moment to introduce ourselves in a cool and relaxed session. Those who did not answer the pre-study questionnaire prior to the meeting also had a chance to fill it out (more details about this survey are available in Section~\ref{sec:data-collection}).

    \item \textbf{\tool Demonstration (2:15 PM – 2:50 PM)}: We conducted an \tool demonstration, emphasizing its key features and ways of interaction. By the conclusion of the demonstration, participants would be equipped to execute a basic ``hello world'' activity using the platform.

    \item \textbf{Exploring \tool (3:00 PM – 4:50 PM)}: With the environment duly set up, participants were engaged in independent exploration. For this exercise, we provided three curated knowledge sources and specific tasks to facilitate hands-on experience. These tasks encompassed various activities: 1) replicating the initial demonstration, 2) employing the given knowledge sources to simulate banking operations like transfers or payments, 3) refining the user's prompt for more effective use of the knowledge sources, and 4) exploring extra knowledge sources relevant to the participant's specific field.

    %\item \textbf{Implementation Phase (4:00 PM – 4:50 PM)}: Up to this point, our focus has been on code generation primarily. However, a crucial step lies in its execution. The question arises: Can we develop a functional product using the generated code? Thus, in this step we aimed to understand to what extend the participants would be able to create a functional code with the \tool output.

    \item \textbf{Group Discussion (5:00 PM – 6:00 PM)}: In this concluding hour, we %seek 
    sought to address and reflect upon their experience while utilizing \tool. We made it clear that the developers' insights and feedback were invaluable in the ongoing refinement of the \tool.
\end{itemize}

A discerning reader might observe that there is a 10-minute break between each session. This interlude was intentionally scheduled to offer participants a chance for stretching, restroom breaks, and other necessities. Moreover, to build rapport and foster engagement with attendees who remained in the room, we initiated discussions about the preceding activity, inquiring about any uncertainties or challenges they may have encountered, for instance, in their machine configuration.

\subsection{Data Collection}\label{sec:data-collection}

We collected data in three different moments. First, we administrated a pre-study survey (Section~\ref{sec:survey});  second, we audio-recorded the conversations that happened during the study (Section~\ref{sec:conversations}); and, third, we conducted a group discussion after the study (Section~\ref{sec:fg}). We will discuss each of these collection methods next.

\subsubsection{Survey}\label{sec:survey}

We established the survey as an online questionnaire.
%, accessible via \texttt{https://zup1.typeform.com/to/B7eWFDb5} (original version in Portuguese). 
Before the official survey release, we piloted the instrument with three practitioners to assess its clarity and relevance. The feedback from these pilot participants allowed us to refine certain queries, ensuring optimal comprehension --- for instance, we adjusted a question about user experience to be more specific based on a suggestion. Following these revisions, the pilot responses were purged to ensure the integrity of the final dataset.

The survey was disseminated company-wide, via our weekly news email, one week before the study. To maximize engagement, the survey was also promoted in various company communication channels. As a pre-requisite to join the session, employees were required to complete the questionnaire. Additionally, at the outset of the study, we emphasized the importance of the survey, allotting the initial 15 minutes for attendees to fill it out. Ultimately, a total of \survey practitioners participated, though pinpointing an exact response rate proved challenging, given the broad outreach juxtaposed with the targeted audience for the study. The survey was crafted using the TypeForm platform; the platform estimated an average completion time of four minutes.

\vspace{0.2cm}
\noindent
\textbf{Questions.}
Our survey had 12 questions (all of them were required and five were open). The survey was not anonymous; in the very first question, we asked for the participant's email. We did so to generate the invite list to the online room, the second phase of the data collection. 
The questions covered in the survey were:

\begin{enumerate}[label=Q\arabic*)]
    \item Enter your email? [Open question]
    \item What is your age? [Open question]
    \item What is your technical profile? [Open question]
    \item How long have you been working with this technical profile? [Numerical scale \{0 to 10\}]
    \item How would you rate your experience in the following programming languages? [Numerical scale \{1 to 5\}], for the following programming languages: Java, C\#, Go, Python, JavaScript, and TypeScript
    \item Have you ever used a Generative AI tool for code generation? [Choices: \{Yes/No\}]
    \item Which Generative AI tool for code generation do you use most frequently? [Multiple Answer: \{Github Copilot, Amazon Whisperer, Sourcegraph Cody, Other\}]
    \item How often do you use this tool? [Numerical scale \{1 to 5\}]
    \item How useful is the output from these tools to you? [Numerical scale \{1 to 5\}]
    \item Do you need to modify the code generated by these tools before making a commit? [Numerical scale \{1 to 5\}]
    \item What features provided by these tools do you find most interesting? [Open question]
    \item What features would you like these tools to implement? [Open question]
\end{enumerate}

The complete set of questions, as well as the actual survey responses, are anonymized and available at the companion website\footnote{To be published upon acceptance.}.

\subsubsection{Recorded conversations}\label{sec:conversations}

All discussions and interactions that took place during the study were recorded, having obtained the explicit consent of the participants. At the welcoming session, we clarified that attendees who might be hesitant about the recording could still actively engage with the tool. However, we requested that they refrain from joining the public discussions through video or text. Instead, they were encouraged to communicate using specified private channels. Notably, we found that every attendee was receptive to the recording procedure, with none opting for the private communication channels. The total duration of the recorded video amounted to 4 hours and 2 minutes.

\subsubsection{Group discussion}\label{sec:fg}

Our group discussion mirrored the one adopted by Luz and colleagues~\cite{Luz:2019:JSS}. The organization of the group discussion was as follows: (1) a researcher-moderator helmed the session, outlining discussion subjects for the participants; (2) as each topic was broached, participants presented their thoughts, and keywords were posted on the shared slides; (3) subsequently, with the notes on the slides, the participants could provide additional comments.

The group discussion happened immediately after the technical session. We used the same Google Meet call to conduct the discussion. However, many practitioners were unable to attend the whole study due to other commitments. Therefore, the discussion started with 34 participants and ended with 20. When we inspected the recorded video, we observed several interactions among the participants, usually complimenting \tool. So, although the group discussion concluded with 20 participants, we actually had around 25 participants engaged in the conversation (not to mention the interactions via chat in the meeting room).

Although we tried to reach different participants in the online room, due to the high number of attendees, not all of them were able to express their perspectives. The group discussion lasted approximately 1 hour and we sought to answer the following questions during this phase:

\begin{itemize}
    \item Did you find any issues with running \tool that halted your progress?
    \item Did you feel the need to understand more about the provided knowledge sources?
    \item How did you perceive the ease of use of \tool? (What factors influenced this evaluation?)
    \item What were the primary benefits you derived from using \tool in your project?
    \item What challenges did you face when using \tool?
    \item How useful and accurate were the responses generated by \tool for your purpose?
    \item What other features would you expect \tool to offer?
    \item Did \tool save you time during the development process? (If yes, how?)
    \item How likely you would be to integrate \tool into your daily work routine?
\end{itemize}

%\begin{itemize}
%    \item Were you able to run the \tool, or did you encounter a bug that halted your progress?
%    \item Were you able to utilize the provided knowledge sources? Did you feel the need to understand more about them?
%    \item How would you rate the ease of use of \tool? What factors influenced this rating?
%    \item What were the primary benefits you derived from using \tool in your project?
%    \item What challenges did you face when using \tool?
%    \item Did you find the responses generated by \tool accurate and useful for your purpose?
%    \item Was there a feature you expected to find in \tool but was absent?
%    \item Do you believe that \tool saved your time during the development process? If yes, in what ways?
%    \item Would you integrate \tool into your daily work routine?
%\end{itemize}

\subsection{Data Analysis}\label{sec:data-analysis}

We employed diverse data analysis methods, according to the data collected. To analyze the survey delineated in Section~\ref{sec:survey}, we used descriptive statistics to provide a concise summary of the primary information. For the open-ended questions, open coding techniques were utilized to classify the answers.

To analyze the recorded conversations and the group discussion (Section~\ref{sec:conversations} and Section~\ref{sec:fg}) we made use of a distinct approach. We developed a software tool that automatically downloaded the video, extracted its audio, and leveraged the OpenAI Whisper model\footnote{https://github.com/openai/whisper} for transcription. This process yielded text data comprising 29,851 words (45,580 tokens\footnote{Computed with Tiktoken library, \texttt{https://github.com/openai/tiktoken}.}). To identify predominant categories and themes within this text, we queried GPT-4. For example, GPT-4 assisted in enumerating the most recurrent questions posed during the meeting and pinpointing prevalent issues highlighted by participants. We then manually refined GPT-4's output, supplementing it with pertinent observations that the model might have overlooked. 
We conducted two approaches as a way to mitigate study hallucination problems\footnote{Hallucinations are common in LLM-based tools. See Bang's et al. discussion of ChatGPT~\cite{Bang2023}.}. First, one author read the full transcript while watching the recorded video; during this task, the author fixed minor errors in the transcript, making it more accurate. Second, we asked two practitioners who joined in the study to analyze the list of categories and themes produced by GPT-4, remove them if they found them wrong, and complement them with additional ones that they found representative (although missing from the initial list). The practitioners mentioned that the categories in the GPT-4 list are accurate and no additional items were provided.

The researchers involved also analyzed the themes and the data, and, although they agreed with GPT-4 classification, they judged that there was some overlap across the categories. After discussions, they came to a consensus on keeping four main categories: 1) general questions, 2) perceived benefits, 3) challenges encountered, and 4) perception of productivity. GPT-4  suggested one other category, called ``usefulness of the generated answer''. We dismissed this last category for the sake of traceability since their themes were following up on those themes from categories 2, 3, and 4. 
We elaborate on each one of these categories throughout Section~\ref{sec:results}.

\subsection{Participants Demographics}

In this section, we present the demographic details of the \survey respondents to our survey. 

\begin{figure}
    \centering
    \fbox{\includegraphics[width=0.46\textwidth]{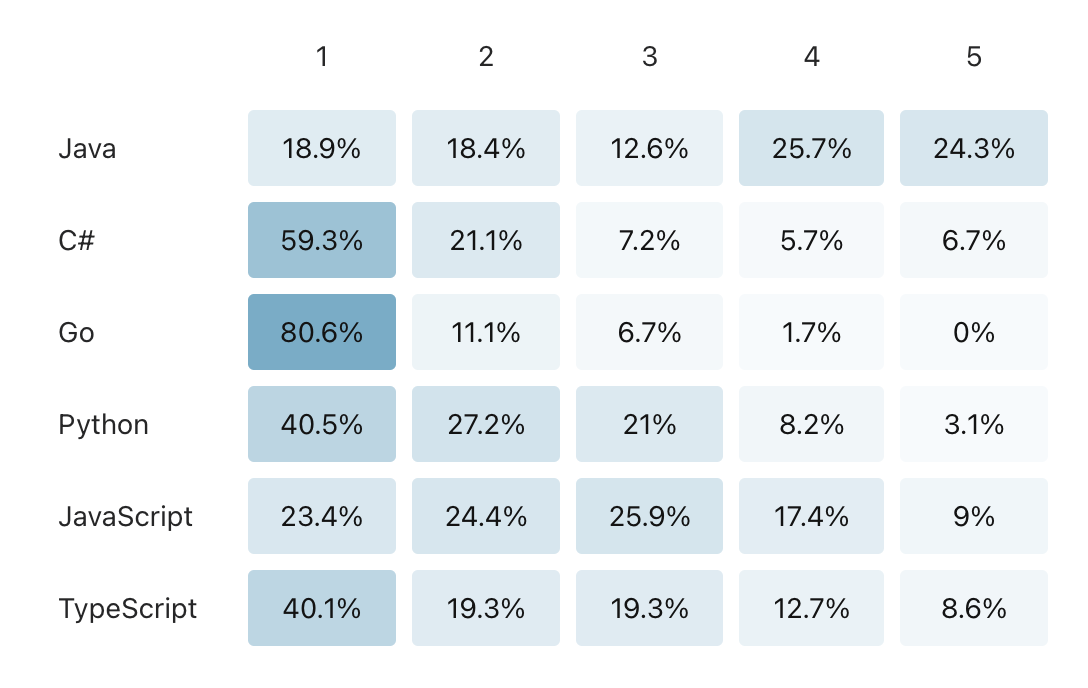}} \\
    \caption{Participant Programming Experience.}
    \label{fig:programming}
\end{figure}

\begin{figure}[h]
    \centering
    \textsc{(a) Use of GenAI Tools}\\
    \fbox{\includegraphics[width=0.46\textwidth]{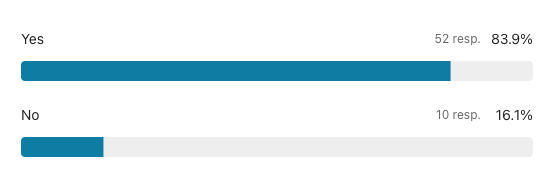}} \\
    \vspace{0.2cm}
    \textsc{(b) Which GenAI Tools}\\
    \fbox{\includegraphics[width=0.46\textwidth]{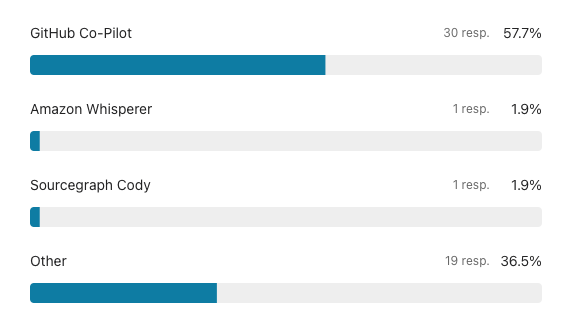}} \\
    \caption{Participants experience with GenAI-based tools.}
    \label{fig:genai-tools}
\end{figure}

The average age of our participants is 34 years, with 6 years of experience in software development, and 2.6 years affiliated with the company. When evaluating proficiency in programming languages (Figure~\ref{fig:programming}), a significant portion (50\%) self-identified as proficient in Java (columns 4 and 5 in Figure~\ref{fig:programming}). This was followed by JavaScript (26.4\%), TypeScript (21.2\%), C\# (12.4\%), Python (11.2\%), and Go (1.7\%). Figure~\ref{fig:genai-tools} shows the percentage of the participants who had experience with GenAI tools (Figure~\ref{fig:genai-tools}.a), and which ones (Figure~\ref{fig:genai-tools}.b). Notably, 83.9\% of the respondents have prior experience with coding AI assistants, with GitHub Copilot being the predominant choice (57.7\%). 

Other AI coding assistants, each mentioned by a single respondent, include Amazon Code Whisperer, Sourcegraph Cody, AskCodi, Codeium, Phind, and ChatGPT (with 11 mentions). A noteworthy observation is that 58.8\% of the participants seldom utilize these tools; in fact, 17.6\% have never employed them. A minority, 23.5\%, incorporate these tools regularly in their workflow. When assessing the utility of the output from these assistants, 29\% of the participants found them useful, and an additional 19.4\% deemed them highly valuable. However, 46.7\% reported that they often make substantial modifications to the outputs these tools produce. The rest, 4.9\% had a neutral answer to this question.

\section{Results}\label{sec:results}

Given the exploratory nature of our study, we organize our results according to the four themes that emerged in our analysis. 

\subsection{General Questions}

We start by describing general questions that the participant had when first interacting with \tool. Understanding general questions is important because developers may need to answer them before they can write any code.

\vspace{0.2cm}
\noindent
\textbf{What are the main features and functionalities of \tool?} Participants expressed interest in understanding how \tool operates and its distinction from other generative AI tools in terms of feature set. They sought detailed insights into the specific functionalities that set \tool apart, including its unique capabilities and how these functionalities enhance the user experience compared to other available tools.

\vspace{0.2cm}
\noindent
\textbf{How to set up and start using \tool?}
This question focused on the setup and initial configuration requirements for \tool. Participants, who were used to other GenAI tools with minimal setup, inquired about a similarly streamlined process for \tool. They were particularly interested in \tool's integration with various IDEs via plugins, concerning support for a wide array of IDEs not yet compatible (such as XCode and Eclipse). Additionally, there was a keen interest in understanding the role and selection of knowledge sources for \tool, including the number and types of sources to use for optimal performance.

\vspace{0.2cm}
\noindent
\textbf{What are the benefits of using \tool compared to other similar tools?}
Participants, experienced with other GenAI tools, frequently asked about \tool's advantages over these tools. A notable inquiry centered on the efficacy of \tool in the absence of user-provided knowledge sources, especially in comparison to tools like CoPilot, which also utilize state-of-the-art OpenAI models. We clarified that without specific knowledge sources, \tool may not demonstrate significant performance enhancements. This discussion underscored that \tool is not a \textit{panacea}; effective use requires users to actively engage in the selection and design of knowledge sources to fully leverage its capabilities.

\vspace{0.2cm}
\noindent
\textbf{Does \tool actually bring productivity gains and time savings for developers?}
Another recurrent general question was about the tangible benefits of \tool in enhancing software development productivity\footnote{Note that we are reporting what our informants said, without discussing whether the concepts they used, e.g., productivity, are accurate or not.}. Participants were keen to learn how \tool translates its features and capabilities into real-world time savings and efficiency improvements for developers. One respondent said that ``\textit{I asked him} [the tool] \textit{to generate unit tests, he did, the tests were good. I said, that's good, but I want another one. He went and did it right too. I saw that he really can generate test and now it's something I'll have to worry less, because I can leave it up to him to do it. I think this will help a lot on a daily basis. It'll speed things up a lot.}''

%\textcolor{red}{Thayssa, quote aqui? ==> COLOQUEI 3 NOS COMENTÁRIOS PRA ESCOLHA} 

\subsection{Perceived Benefits} \label{sec:benefits}

In this section, we delve into the potential contributions of \tool in simplifying and enhancing the coding process. The participants' experiences, initially tinged with skepticism, evolved into recognition of \tool's capabilities in generating precise code for complex API integrations. We explore how \tool's contextual understanding, quick commands, and interactive code refinement through chat significantly streamline the development cycle.

\vspace{0.2cm}
\noindent
\textbf{Generation of accurate codes for swift integration with internal APIs}
%\textbf{Generation of accurate and complete codes for quick integration with internal APIs.} 
Initially, participants expressed skepticism about the accuracy and completeness of the code generated by \tool, particularly for complex API integration. During the study, the participants were handed a few API files, which they used to generate integration using \tool.
Their experience revealed \tool's ability to generate functional code snippets, acknowledging this as a crucial feature for accelerating development cycles and reducing errors. This was highlighted by P4, who mentioned the following:
``\textit{Thus, the tool accurately grasped the context, generating a code with Spring framework --- it was impressive. I did not expect it to work out so well. \tool created the class as I requested. Indeed, I am quite amazed here.}''
%\textcolor{red}{Thayssa, quote aqui?}.

\vspace{0.2cm}
\noindent
\textbf{Faster and more context-aware responses and code suggestions compared to other tools.} 
Participants initially perceived \tool as an enhancement to general-purpose AI coding assistants, intrigued by its potential for quick, context-sensitive solutions. Upon further use, they recognized \tool's significant value in providing time-efficient, relevant, and precise coding suggestions specifically tailored to their project's context. As participant P5 mentioned: 
``\textit{For me, the contextualization helped in generating accurate code. It helped a lot. I saved time; the answer was straight to the point. The contextualization was the biggest gain for me}''.
%\textcolor{red}{Thayssa, quote aqui?}

\vspace{0.2cm}
\noindent
\textbf{Iterative refinement of generated codes through chat interaction.} 
Given that \tool operates as a conversational agent equipped with an internal memory to record past interactions, users can engage with it in a manner akin to a chatbot. This feature of maintaining a history of previous conversations was initially viewed as interesting. It facilitates interactive code refinement and assists in honing code outputs to align precisely with specific project requirements, thereby enabling users to achieve more optimal coding solutions.

\vspace{0.2cm}
\noindent
\textbf{Support for repetitive and routine tasks through ``quick commands''.} 
Quick commands are shortcuts offered by \tool, which developers could use to automate common software engineering tasks, such as creating tests, documenting code, or even asking the tool to explain a certain code snippet. 
Participants initially underestimated the impact of this feature, considering it a minor convenience for routine coding tasks. As they became more familiar with \tool, the collective sentiment shifted to view these quick commands as time-savers, greatly aiding in automating mundane aspects of coding and allowing them to focus on more complex tasks, assisting in improving code reliability and maintenance. One participant acknowledged the use of ``quick commands'' as a potential enhancement to their development workflow.

\vspace{0.2cm}
\noindent
\textbf{Time-saving by centralizing information access.}
%\textbf{Time-saving by not having to search for information in different places.} 
Participants initially recognized %the feature as a tool for convenience, 
\tool as a convenience by minimizing the need to alternate between various information sources. With continued use of \tool, they appreciated its efficiency in providing centralized access to essential information within the IDE, notably reducing development time and cognitive load. This was underscored by participant P1: 
``\textit{I obtained the answer from the code \tool generated; I didn't need to go to the original knowledge source. In this case, it was accurate and helpful. [...] Even though the question I asked was a very simple example, \tool demonstrated that centralizing information in the IDE would be highly useful}.''
%\textcolor{red}{Thayssa, quote aqui?}.

%\vspace{0.2cm}
%\noindent
%\textbf{Easier access to documentation and guidelines within the Integrated Development Environment (IDE) setting.} 
%At first, participants viewed this as a minor enhancement, simplifying the access to documentation. But upon using \tool, they understood the substantial benefit of having immediate access to relevant guidelines and documentation within their workflow, significantly enhancing their productivity and reducing interruptions in their coding process.

%\vspace{0.2cm}
%\noindent
%\textbf{Ability to automatically generate tests and documentation.} The automatic generation of tests and documentation was initially met with curiosity but some skepticism regarding the quality and relevance. As participants used \tool, they realized the immense potential of this feature in ensuring code reliability and maintaining project documentation, which they acknowledged as a significant enhancement to their development workflow.

\subsection{Perception of productivity} 

%After discussing the potential benefits and challenges, we asked the participants' perception of whether \tool would be able to reduce their programming times.

After discussing the potential benefits of using \tool, we asked participants' opinions about its potential impact on their productivity.

\vspace{0.2cm}
\noindent
\textbf{Contextualized code snippets.} A few participants affirmed that \tool indeed saved their time by providing quick and contextual code snippets during the study. The ability of \tool to quickly understand the context and deliver precise code snippets not only streamlined the coding process but also allowed users to focus more on creative and complex aspects of their work. When asked one participant mentioned: \textit{``Yes, because within the context of each project, the knowledge base will become increasingly richer and will improve in generating responses.''}

\vspace{0.2cm}
\noindent
\textbf{Aggregating knowledge sources within the IDE.} As mentioned among the benefits, participants specifically mentioned that \tool saved time by eliminating the need to search for knowledge sources because they are now ``available'' within the IDE. This feature was highlighted as a major time-saver, therefore influencing the perception of productivity of some study participants. For instance, one respondent mentioned: \textit{``It} \tool \textit{``saves time, as it's a shortcut for accessing information. If done through conventional means, you'd have to search through a search engine, consult books, or find people with that information to assist you, which would certainly take longer.''} Another participant mentioned that by providing knowledge sources within the IDE, \tool could also reduce interruptions: ``\textit{I believe it can save our time, as it allows us to reduce interruptions when seeking specific information.}'' In summary, by providing instant access to relevant knowledge sources directly within the tool, \tool enabled users to access necessary information or code samples without disrupting their workflow.

%- The mention of having pre-configured settings (like workspaces, stacks, knowledge sources) was highlighted as a feature that could bring even greater agility without needing to delve into the technical details. This would allow users to quickly start working on their projects with an optimized environment, tailored to their specific needs, and would be particularly beneficial for beginners or those unfamiliar with setting up complex development environments.

\vspace{0.2cm}
\noindent
\textbf{Productivity gains unlock \textit{only if} users know how to use \tool.} A participant commented that, like any AI tool, \tool only brings time-saving benefits if used correctly with refined prompts and proper settings. If used incorrectly, it could even lead to time wastage. For instance, one participant complemented the following: ``\textit{However, if used carelessly or by less experienced people, it may result in more work for the more experienced developers.}''
%``Porém se usado de maneira descuidada ou por pessoas menos experientes poderá acarretar em mais trabalho para os desenvolvedores mais experientes.''
This insight underscores the importance of understanding how to effectively interact with AI tools. Properly formulated queries and a clear understanding of \tool's capabilities are essential to harness its full potential and avoid counterproductive outcomes.

\vspace{0.2cm}
\noindent
\textbf{Insufficient experience to evaluate.}
Finally, a few participants mentioned that the participation in the study was not enough to draw a definitive perception of the productivity gains of \tool, as one engineer highlighted: ``I don't think I used it for enough time and in scenarios that would allow me to answer this question.''
%``Não acho que utilizei tempo e em cenários suficientes pra responder essa pergunta''.

%\subsection{Unexpected Outcomes} 
\subsection{Challenges Encountered} 
\label{sec:challenges}

This section highlights the various challenges %, technical limitations, and user experience aspects 
encountered during the participants' interaction with \tool. We divided these challenges into three groups: (i) challenges associated with the adoption of the Retrieval-Augmented Generation (RAG) technique, specifically the knowledge sources used; (ii) challenges associated with large language models in general; and, finally, (iii) other technical and user challenges associated with either UI aspects or user expectations. We present the challenges according to these groups.

\vspace{0.2cm}
\noindent
\textbf{Figuring out what is a good knowledge source.} 
As mentioned in Section \ref{sec:approach}, knowledge sources are representative documents that enrich the prompts for RAG's generation component, providing essential context for task development. Without these sources, responses from \tool would be less contextualized, resembling the answers from general-purpose coding AI assistants. Thus, identifying effective knowledge sources is vital for \tool's performance. Our study, however, revealed that not all participants were able to understand what constitutes a (good) knowledge source. This was observed during group discussion about other kinds of knowledge sources they would use, based on their team context. While a few participants were able to give interesting examples (e.g., using a database schema as a knowledge source, and asking \tool to create SQL queries based on it), other participants were unable to come up with one single example. Furthermore, others gave examples that were not based on coding tasks, and a few participants even mentioned that they ``\textit{need to understand more about it, but were able to use it in a very basic way.}''
%``Preciso entender mais sobre, mas consegui utilizar de uma forma bem inicial''.

%\vspace{0.2cm}
%\noindent
%\textbf{Impact of Mixing Knowledge Sources.} The mixing of different knowledge sources was perceived as affecting the precision of the code suggestions. Users found that the blending of information sometimes led to less accurate or relevant code suggestions, highlighting the need for better source management. \gnote{explicar melhor aqui} \textcolor{red}{Thayssa, teria quote aqui também?}

\vspace{0.2cm}
\noindent
\textbf{Knowledge Source Mixing Impact.} During the focus group, it was mentioned that mixing several different knowledge sources affected the accuracy of the responses in certain cases, i.e., users found that the blending of information sometimes led to less accurate or relevant code suggestions, highlighting the need for better source management. This observation suggests that while having access to a wide range of sources can be beneficial, it also poses a challenge in ensuring that the information drawn from these sources is relevant and accurately integrated.
One participant highlighted this issue as the following: ``\textit{Initially, I thought using various Knowledge Sources in the same workspace might not be a good practice. This led me to experiment with my own project. I combined a postal code API and various elements from the provided knowledge source. The result wasn't great}.''

%\textcolor{red}{thayssa, them uma quote aqui?} 

\vspace{0.2cm}
\noindent
\textbf{Inacurrate code suggestions and prompt refinement.} Participants noted the necessity to refine and adjust prompts to obtain accurate responses. For instance, one participant mentioned that ``\textit{In some cases, they} [the code suggestions] \textit{were not accurate; in others, they required many interactions and didn't yield the expected result.}'' Another participant added that ``\textit{It wasn't as accurate; in my case when I entered it as 'Go'} [the programming language] \textit{it generated generic things. It seems that it works better in Java.}'', revealing a potential bias towards more popular programming languages. This feedback underscores the importance of clear prompt formulation when interacting with AI tools. It also points to the potential need for iterative interaction, where initial responses serve as a starting point for further refinement to achieve the desired outcome. Furthermore, this result highlights how the participants required additional effort and understanding of how to effectively communicate with the AI, which was a learning experience for several participants. %\textcolor{red}{Thayssa, uma quote aqui serial legal}. 

\vspace{0.2cm}
\noindent
\textbf{Response Inconsistency:} Our informants revealed that \tool sometimes provided varied responses to identical prompts. This inconsistency in output led to confusion among users and raised questions about the reliability of the tool in repetitive tasks. Although in the LLM literature, it is well-discussed that slightly different prompts could lead to different answers, users found this experience awkward, potentially negatively impacting their trust in \tool.

\vspace{0.2cm}
\noindent
\textbf{Generating complex code structures.} \tool faced difficulties in suggesting complex code like ready-to-use controllers. %, proved difficult for \tool. 
Participants expressed that while basic code generation was effective, the tool struggled with more sophisticated coding requirements. For example, one participant expressed this saying ``\textit{I tried to use the stack based on Spring, Java, Kotlin, and then play with the knowledge sources. However, despite my efforts, \tool was unable to generate the code with the endpoints. Instead of using controllers, it created methods in a main class that starts the SpringBoot app.}''

%\textcolor{red}{Thayssa, quote aqui.}.
%Moreover, generating complex code structures, like ready-to-use controllers, proved difficult for \tool. Participants expressed that while basic code generation was effective, the tool struggled with more sophisticated coding requirements.

\vspace{0.2cm}
\noindent
\textbf{Inability to deal with custom languages.} A participant proposes the ability to add custom language support for code snippets. The participant mentioned: ``\textit{When trying to include a snippet with a Cucumber Gherkin code, Gherkin does not appear in the list that defines what type of language the code refers to. Is it possible to register it?}'' This may limit the ability to deal with specific/custom language and impact the outcomes for specific projects.

%\vspace{0.2cm}
%\noindent
%\textbf{Context Size Limitations:} A notable limitation encountered by participants was the constraint on the size of the context that could be sent to the AI. This limitation restricted the depth and breadth of interactions, especially in complex coding scenarios.

\vspace{0.2cm}
\noindent
\textbf{Conversation History Loss.} A common frustration among participants was the loss of conversation history upon closing the IDE. This issue was particularly problematic for those working on complex tasks over extended periods, as it disrupted the continuity of their work and thought processes.

\vspace{0.2cm}
\noindent
\textbf{Missing User-Expected Features:} Participants highlighted the absence of certain functionalities in \tool that they had anticipated. This gap in expectations versus reality suggested a misalignment between the tool's capabilities and the users' needs. Similarly, the absence of adequate plugin availability and support for other IDEs was mentioned as a drawback, which restricted the usability of \tool across different development environments, impacting its adaptability.

\vspace{0.2cm}
\noindent
\textbf{Initial Configuration and Onboarding Process.} At the beginning of the study, participants struggled with the setup process of \tool, finding the integration of elements like workspaces, AI stacks, and knowledge sources challenging. This initial complexity was a significant barrier for many---particularly for those less experienced with such environments---indicating a need for a streamlined onboarding process. Indeed, a few participants were unable to correctly set up the environment, and thus did not actively participate in the full study. Still, one participant mentioned that ``[the code suggestions] \textit{could be better, but I believe it was because of how I configured the Knowledge Sources.}'' This feedback shows that \tool's effectiveness is significantly related to appropriately configuring the environment.

\vspace{0.2cm}
\noindent
\textbf{Technical Limitations.} Participants frequently encountered technical issues such as timeouts and error messages (403 and 500 statuses). Additionally, some reported accessibility problems on specific machines (``\textit{I couldn't use \tool with the client's machine. Does this mean that I was only able to run \tool on my personal computer?} - P6).
%\textcolor{red}{Thayssa, citar 1 ou 2 exemplos aqui. nao precisa ser quotes}. 
Another limitation mentioned was the impossibility of having a Git repository as a knowledge source hinders the ability to understand and work with new or existing projects, with all information available in their Git repositories. In this context, a participant stated that: ``\textit{The possibility to add a complete Git repository as a knowledge source. It would help A LOT in adding a quick context of documentation or an application.}''%It would also be important to create client-side hooks that would automatically feed the model every time a new commit is merged into this repository. This would avoid challenges related to deprecated information, and keep a good accuracy when producing new pieces of code.
These technical limitations hindered the smooth operation of \tool, affecting the overall user experience.

\section{Discussion and Future Opportunities}
\label{sec:discussion}
Reflecting on our results, we noticed that using a contextualized model to support developers will be beneficial for the company developers. This was evidenced by the benefits listed in Section~\ref{sec:benefits}. 

\vspace{0.2cm}
\noindent
\textbf{On the Accurateness of Code Generation.}
Several participants highlighted that the code suggestions they received from \tool were both useful and accurate for their purposes, enabling them to simply copy and paste the generated code directly into their projects. For instance, a developer was able to quickly generate an integration with an internal API of a specific client by simply requesting it through \tool. Being able to use the provided code without additional modifications not only saved time but also demonstrated the tool's capability to understand and address specific coding needs accurately. This exemplifies the potential time-saving benefits of the tool, showcasing its ability to automate and simplify complex tasks. This is in line with other studies~\cite{vaithilingam2022expectation, Bird2023, Anonymous}, adding evidence related to the power of models fine-tuned for specific contexts.
Participants also noted that some suggestions were not entirely precise, necessitating adjustments and refinement of the prompts. Refining prompts is a strategy used by other software developers using academic (see ~\cite{Xu2022}) or proprietary tools (see ~\cite{Anonymous}). This is a challenge associated with LLMs in general, but, as we discussed in section \ref{sec:challenges}, we also identified challenges related to the usage of knowledge sources and other technical and user-related challenges.

\vspace{0.2cm}
\noindent
\textbf{Usability and Developer Efficiency.}
From a usability perspective, important results are highlighted. First, the ability to offer shortcuts (quick commands) has been shown to be beneficial to the developers, supporting what Barke et al. called the acceleration mode~\cite{barke2023grounded}. Our results indicate that the available tool did support that working mode. Second, \tool's design, based on an interactive chat instead of focusing on code completion~\cite{barke2023grounded}, also provided an opportunity for the iterative refinement of code suggestions as well as avoided interruptions~\cite{Bird2023, Anonymous} similar to what has been observed by Ross et al.~\cite{Ross2023}. Finally, the possibility to focus solely on the IDE while seeking information has been shown to be important for software developers using AI coding assistants~\cite{Anonymous}.

\vspace{0.2cm}
\noindent
\textbf{Having the right mix of knowledge sources is important.}
One important challenge that we observed is associated with knowledge sources. As mentioned in section \ref{sec:approach}, these sources are used in the RAG approach to enhance LLM-generated content by anchoring it in external knowledge sources. Therefore, it is not surprising that our informants reported different aspects associated with it including the difficulty in identifying good knowledge sources, 
the importance of properly configuring \tool with the knowledge sources, and finally the negative impact of mixing different sources. An interesting research avenue would be exploring the efficiency of different knowledge sources, and creating systematic approaches to do so, supporting (semi-)automatic ways of optimizing the creation of contextualized coding assistants.

%Another challenge associated with LLMs in general was the one related to the inability of \tool to work with custom languages. %This becomes important to find ways to make integration with custom languages available. In the example given by the participant, they mentioned Gherkin, however, such a feature would allow users to register any language not currently available in the predefined list. This enhancement aims to ensure proper formatting and recognition for a wider range of programming and scripting languages, enhancing the tool's flexibility and inclusivity. This inclusion may entail, for example, finding out the correct sources that could provide the information required to generate the correct outcome. One potential line for development and research is to add a model that would support discovery in the universe of available knowledge sources that could be available, to provide an accurate answer. Another possibility, more reactive, would be asking the users to help by providing new sources to feed the model with information about that specific language.

\vspace{0.2cm}
\noindent
\textbf{Challenges and Improvements in LLMs.}
We also highlighted several perceived challenges using \tool. It was interesting to find out that participants reported that the tool was able to generate accurate code snippets, that helped to neatly integrate internal APIs. Our study showed that it is possible to reduce the issues with incorrectness~\cite{Nguyen2022, Yetistiren2022} by creating a contextualized model to retrieve information from contextual knowledge sources. In contrast, the non-deterministic nature of LLMs~\cite{Ouyang2023} remains an issue, i.e., some of the challenges are inherent to LLMs, while other challenges are purely technical (conversation history loss, initial configuration, etc) and will be addressed in future releases.

\vspace{0.2cm}
\noindent
\textbf{Enhanced Generic Configuration Options for AI Coding Assistant. }
Since participants reported challenges with the initial configuration, one potential way to move ahead would be providing a more flexible and generic way of configuring \tool.
The goal here is to make the setup more user-oriented, without the need to include a lot of information before the use of the tool. %As one participant mentioned, ``\textit{the 'Definition' part of the \tool could be more generic, an open text field instead of the various options. It would be nice to even inform coding style preferences (e.g., using code-guards, methods with JS Doc, preference for more explanatory identifiers, etc).}'' 
This comes with a trade-off, since when the settings are properly configured, the responses are very precise and save time---therefore, making it too flexible may hinder less experienced team members.
The setup was idealized in a way to get the key configuration/customization items from the user, to make the code suggestions more reliable and precise. By making it less constrained, we may affect the accuracy of results. More investigation is required to understand how much flexibility is possible, without negatively impacting the outcome.

\section{Limitations}
\label{sec:limitations}
This study, while extensive, has notable limitations. Firstly, our data collection involved a sizeable sample of practitioners using a contextualized coding assistant for the first time. However, this sample may not fully represent the broader population of software developers. Still, although the majority of the participants have previous experience with Generative AI tools, such as ChatGPT and GitHub Copilot, their experience with these tools might not naturally translate to the use of \tool, in particular, because they have to select and design representative knowledge sources, which is, by design, an important effort --- which is not required by general-purpose tools.

Secondly, due to company policy recommendations against requesting gender information during the prestudy to avoid participant discomfort, our study did not gather this data. This omission restricts our ability to conduct comprehensive comparative analyses across different gender groups.
Another limitation concerns the robustness of \tool, currently in its beta phase. Some challenges noted by participants might stem from insufficient testing rather than flaws in the underlying Large Language Model (LLM). This factor could adversely affect the overall user experience.

Given the nature of how the group discussion was conducted (as an online call with dozens of participants), we were unable to accurately identify the number of participants who mentioned a given benefit/limitation. As such, during our discussion section, we often refer to them as `many', `a few', `various', and the like.

Moreover, our data analysis partially relied on AI-based tools. While these tools excel in summarization tasks~\cite{miller2019leveraging}, they do not adhere to stringent qualitative research methodologies. Consequently, while we successfully extracted representative quotes, the analysis's rigor cannot be fully assured. To address this, two authors involved in the study reviewed and validated the AI-generated outputs. These authors suggested no additional items.

\section{Conclusion}
\label{sec:conclusion}
The recent flow of newly introduced AI coding assistants has unlocked developers' potential, in a myriad of coding tasks. However, these coding assistants, when not trained taking into account the developers' context (with their representative documents, coding styles, etc), might produce answers that although appear interesting at first, may not be as precise as developers' need. Both academia and industry have recognized the need for more intuitive, conversational AI tools that could seamlessly integrate with existing IDEs, providing real-time, context-aware assistance. 

In this work, we explored the use of \tool, a conversational AI tool, which is enriched with developers' representative documents to generate more appropriate answers. We used the tool in a controlled setting with \survey practitioners.
Our findings revealed that \tool could improve productivity and time efficiency. Participants appreciated its ability to quickly generate accurate code snippets and contextual code suggestions. However, the effectiveness of \tool was contingent on precise prompt formulation and optimal configuration of knowledge sources. Some challenges, like technical limitations and the need for better support in diverse IDE environments, were identified, highlighting areas for improvement. Other challenges are inherent to LLMs and require further AI research.

\subsection{Future work}

In future work, we plan to expand the scope of our research activities based on the insights gained from this paper. A key area of exploration will be to conduct longitudinal studies with developers who use \tool over extended periods. This will provide deeper insights into how prolonged use affects productivity, learning curves, and code quality. This will also allow us to understand how such a tool will impact developers' work practices.  Additionally, comparative studies involving other AI-assisted coding tools will offer a broader perspective on \tool's unique strengths and areas for improvement. We also aim to investigate how \tool operates on a wider range of programming languages, assessing its adaptability and effectiveness across diverse coding scenarios. Further, exploring the impact of \tool on team dynamics and collaborative coding practices could provide valuable insights into its role in team-based development settings. Finally, delving into user customization and personalization aspects of \tool could reveal how tailored experiences influence developer satisfaction and tool efficiency. These research activities will collectively contribute to a more comprehensive understanding of AI-assisted coding tools in software development.

\subsection{Artifacts Availability}

This research used two main sources of data: survey responses and audio transcripts from the study sessions. Upon acceptance, these materials will be made available, with a focus on maintaining participant anonymity and adhering to company policies. This ensures a balance between data transparency and ethical considerations in research involving human participants.

\bibliographystyle{ACM-Reference-Format}
\bibliography{references}

\end{document}